\documentclass[10pt]{avicps} % Use file avicps.cls
\usepackage{amsmath}
\usepackage{ae}                % Enables non-english letters, such as ���.
\usepackage{times}             % Using the times package, so that LaTeX and MS Word
\usepackage{url}               % Handling of URL strings.
\usepackage[pdftex]{graphicx}   % Include graphics.
\graphicspath{{pdf/}}

\usepackage[caption=false,font=footnotesize]{subfig}
\usepackage{url}

\begin{document}

\title{Hardware Software Co-design for Automotive CPS using Architecture Analysis and Design Language}

% Use authorinfo, not normal \author
\authorinfo
{Yuchen Zhou ${}^1$ \quad John Baras${}^1$  \quad Shige Wang${}^2$}
{${}^1$The Institute for Systems Research and Electrical and Computer Engineering Department, University of Maryland, College Park, Maryland, USA, $\texttt{\{yzh89,baras\}@umd.edu}$\\
 ${}^2$Electrical and Controls Systems Research Lab,General Motors Research and Development, Warren, Michigan, USA $\texttt{shige.wang@gm.com}$}

\maketitle
\begin{abstract}
Modern cyber-physical systems (CPS) have a close inter-dependence between software and physical components. Automotive embedded systems are typical CPS, as physical chips, sensors and actuators are physical components and software embedded within are the cyber components.
The current stage of embedded systems is highly complex in architecture design for both software and hardware. It is common in industrial practice that high level control algorithm development and low level code implementation on hardware platforms are developed separately with limited shared information. However, software code and hardware architecture become closely related with the increasing complexity. Correlated requirements and dependencies between hardware and software are emerging problems of industrial practice. We demonstrate in this paper a method to link model based system design with real-time simulations and analysis of the architecture model. This allows hardware software co-design and thus early selection of hardware architecture.
\end{abstract}

% Supply appropriate keywords for your article.

%--------------------------------------------------------------------------------
\section{Introduction}

%automotive embedded systems are typical CPS
Cyber-Physical Systems (CPS) are engineered systems constructed as networked interactions of physical and computational (cyber) components. In automotive industry, the
embedded chips within the cars interact with other chips and the physical parts of the
car. The algorithm within the processors are the cyber components, whereas the actuators,
sensors, buses and physical processors are the physical components. The cyber and physical components are normally modeled differently. One particular view of the system that captures
both the cyber and physical parts of the system is the architecture of the embedded system.

% Architecture design is crucial - why need architecture
Modern embedded systems for control applications, such as those for 
engine control and active safety in automobiles, 
are highly complex and integrated. 
Following system engineering principles, the subsystem of 
each control functionality is usually designed and implemented in isolation
by different groups or organizations
according to the specifications derived from system decomposition, 
and is integrated at a later stage. 
Moreover, control algorithm development at a high, abstract level
and software implementation of the algorithm on a hardware platform
are typically performed separately with limited cross-disciplinary knowledge.  
To achieve high integrity of such embedded systems, % with high integrity, 
architecture design, including both software architecture for control implementation
and platform architecture of hardware and supporting software, 
becomes crucial.
A good architecture not only ensures correct integrated behaviors
of the final system but also enables fast and low-cost integration. 

% Challenges in capturing architecture 
A major challenge in architecture design is to balance multiple 
constraints, including schedulablity, performance, timing, and cost. 
For embedded systems, these constraints usually conflict, and the criteria
used to assess an architecture vary in different domains and organizations. 
As an example, an architecture running each control on a separate
processor, known as \emph{federated architecture}, makes it easier to schedule
the software executing on it compared with an architecture consolidating
multiple controls  on the same processor, known as \emph{integrated architecture}.
On the other hand, the former architecture usually incurs higher hardware 
cost than the latter. 
Since architecture design of embedded systems involves comparisons and analysis
of decisions along multiple, often conflicting, dimensions,  
a methodology that supports architecture modeling and analysis is
therefore essential. 

% Current practices and issues - no quantitative, no analysability
In current industry practices, the architecture of an embedded system
is captured in a very abstract manner
using generic representations such as block diagrams,  
where the constructs for architecture elements are modeled as 
indistinguishable blocks with annotations. 
Detail and quantitative properties of these elements are either 
not captured or captured in a different environment with implicit 
connections to the blocks. 
With such a representation, an architecture model captures only 
the structure of a system that cannot be used directly for analysis. 
As such, the analysis is typically very tool-dependent, which not only
limits the engineering processes where the tool applies 
but also incurs high engineering cost to transfer the information,
interpret the results, and maintain the design artifacts. 
Furthermore, some infrastructure services in supporting software  
(triggers, scheduling policies, tasks and interrupt services, for example) 
and the system parafunctional properties (execution times, 
memory footprint, and I/O delays, for example), 
which are critical for architecture design, 
are assumed to be either captured with the software model 
or implicitly specified in implementation. 
%Lack of accurate and sufficiently-detailed architecture model introduces
%difficulties in communicating with multiple development teams 
%acquiring and validating components from various sources.  
%Integration and system-level analysis of the individually-built components,
%especially when involving those across domains such as software signals from/to
%I/O devices and tasks running on a controller, also become challenging. 
As a result, the architecture for embedded systems is usually over-designed,
resulting in unnecessary high system cost. 

% Our poposed solution and contribution
To improve the embedded system architecture, we develop a method based on hardware-software codesign principles. 
Hardware-software codesign uses unified representations for hardware and software
to exploit the trade-offs between hardware and software to achieve system-level
objectives. 
As such the hardware and software can be designed in parallel while
system-level properties are still analyzable through the development and
are preserved after integration. 
Our method uses a standard Architecture Analysis and Design Language (AADL)
to capture high-level hardware models and software partitions. 
Analysis of the system-level behaviors with integrated software partitions 
on the target hardware can then be performed and adjusted with different levels
of abstraction. 
As the development processes of both hardware and software proceed, 
system-level properties can be validated with more implementation details.
We have applied the method to an architecture design for an adaptive cruise 
control (ACC) system as a case study.  
As an example, we have performed design and analysis, specifically 
for schedulability and data synchronization, for the ACC running
on single core or multicore controllers. 
%timing analysis
The method allows us to systematically perform early verification at high level,
and real-time simulation at implementation level under the same architecture framework. 
The rest of the paper is organized as follows. Section \ref{sec:related} summarizes related projects and research. The remaining sections demonstrate our proposed method on an Adaptive Cruise Control (ACC) development problem. Section \ref{sec:AADL} describes the AADL modeling language and the design process using AADL. Section \ref{sec:ACC} describes the ACC algorithm example with its high level architecture layout. The coarse AADL model created is refined further in section \ref{sec:codesign} when the software group creates a detailed behavior model in Simulink and the hardware group proposes some choices of architectures. The last section describes real-time simulation and timing analysis results based on the architecture model, which lead to a conclusion for hardware architecture selection.

\section{Related Work}\label{sec:related}
%
%\begin{figure*}[!t]%trim=0cm 0cm 4cm 1cm, clip=true,
%\centering
%\subfloat[]{\includegraphics[trim=0cm 0cm 0cm 0cm, clip=true, width=0.55\textwidth]{RCMeta2}
 %\label{fig:RCMeta2}
 %}
%\hfill
%\subfloat[]{\includegraphics[trim=0cm -6cm 0cm 0cm, clip=true,width=0.4\textwidth]{RCMeta1}
 %\label{fig:RCMeta1}
%}
%\caption{(a) shows the SysML Internal Block Diagram of a cruise control example. (b) is the corresponding code generated with some modification. RC Meta is a tool composed of plugin for Enterprise Architecture and  Open Source AADL Tool Environment (OSATE). The software is open source which is easy to adapted to other SysML tools.}
%\label{fig:RCMeta}%
%\end{figure*}

%\begin{figure*}[!t]%trim=0cm 0cm 4cm 1cm, clip=true,
%\centering
%\includegraphics[trim=0cm 0cm 0cm 0cm, clip=true,width=0.7\textwidth]{SysMLbbd}
%\caption{Block Definition Diagram in Rhapsody shown aboves gives dependency and overall relationship of the different blocks.}%
%\label{fig:bbd}%
%\end{figure*}

The existing research related to AADL simulation and analysis can be categorized into two major methods. One is based on the Behavior Annex extension of AADL. B. Berthomieu et al. \cite{Berthomieu09} transform the AADL behavior model to the Fiacre language which is then verified using a Time Petri Net Analyzer (Tina). H. Liu and David P. Gluch \cite{Liu09} translate the AADL behavior model to timed automata and verify in UPPAAL, a formal model checker for timed automata. I. Malavolta et al. \cite{Malavolta09} integrate the AADL toolset OSATE with DULLY, a framework which allows automated model transformation. This allows them to model check the AADL model with the behavior annex using a Labelled Transition System Analyser (LTSA). Since AADL does not have the representation power of many existing behavior modeling and simulation engines such as Simulink, our approach clearly departs from previous approaches and only captures a limited behavior model in AADL. In this paper, AADL only acts as high level guidance during the behavior modeling and design phase. The Simulink model can be exported back to AADL using the method proposed in \cite{Passarini12}. The other methods are based on XML output of the AADL parser in OSATE. The scheduling and memory analysis tool Cheddar, which is XML based, has an AADL importer that interprets the AADL model and performs timing analysis upon it \cite{Singhoff05}. Using a metamodel transformation method, F. Mallet et al. \cite{Mallet09} propose a method to transform an AADL model to MARTE, a UML profile for real-time embedded systems, which can be further verified using a time model in MARTE and Clock Constraint Specification Language. However, these analysis tools do not address the controller performance effects of the schedule, and thus cannot conclude further on the hardware selection problem of embedded systems. Cheddar is used here for early verification and schedulabilty analysis, but the hardware selection conclusion is based on real-time simulation results for the overall system performance.

Other related research is from the perspective of ADLs and hardware-software co-design. One is the MARTE project, a UML profile which has been mentioned earlier. The integration of MARTE with EAST-ADL AUTOSAR proposed in \cite{Espinoza09} gives a way to capture the system engineering process from the requirement model to real-time implementation in AUTOSAR. However, such framework lacks the analysis power provided by AADL. Another project is SysML based modeling, architecture, exploration, simulation and synthesis for complex embedded systems (SATURN), in which a combination of SysML, MARTE and SystemC hardware simulation is used to co-simulate software and hardware components \cite{saturn11}. This is closest to our work. The major difference is that the simulations they perform are at much lower implementation levels. The Simulink simulations we perform are high level real-time simulations of the AADL model which are useful for early stage validation and verification to reduce design cost. 
%AADS is another SystemC simulation engine for AADL model. In \cite{Varona12} the authors proposed a solution to the software-hardware partition problem by evaluating Worst Case Execution Time (WCET) using AADS. In our project, the software partition is developed based on the AADL model and tested on an existing processor to get the estimated WCET instead of simulating the WCET in SystemC.

To conclude, our approach emphasizes early verification and schedulabilty analysis using architecture model, as well as high level real-time simulation to verify performance requirements. Previous approaches either lack the simulation power of the behavior model, or the analysis power to perform timing analysis and resource allocation for architecture model. 

%Main part:
%1. AADL language and ACC system model
\section{AADL Language and Design Process}\label{sec:AADL}

Architecture Description Language (ADL) is designed to capture the software and hardware architecture of an embedded system so as to bridge the gaps of feature design in higher levels and software implementation in lower levels. Because of the closer link between embedded code and hardware platforms in automotive applications, ADL is critical to the design process. Architecture Analysis and Design Language (AADL) is emphasized in this paper as an example of ADL. The language has formal textual and graphic representation, which enables easy integration with UML based languages such as SysML, EAST-ADL and AUTOSAR. The textual description makes it extendable and adoptable to industry practices. Several standard annexes to the language have been developed to extend the scope of AADL in analysis areas. Error Annex defines features to enable specifications of redundancy management and risk mitigation methods such as safety, reliability and integrity. Behavior Annex describes the system behaviors as automata and enables formal model checking and code generation based on the AADL model. The open source tool OSATE developed by Peter Feiler's group from SEI CMU, provides parser and model generation as well as basic analysis tools. AADL is chosen also because the OSATE toolset has been demonstrated in several projects in Europe and US, including System Architecture Virtual Integration (SAVI) from the Aerospace Vehicle Systems Institute and Correctness and Modeling and Performance of Aerospace Systems (COMPASS).%\cite{AADLProj}
Comparing to EAST-ADL and other ADLs available, AADL is much more mature and suitable for industrial practice.  Because of the direct linkage between EAST-ADL and AUTOSAR, EAST-ADL is preferred for automotive embedded systems, in particular when the tool chain includes much more powerful analysis or provides far better interfaces for other analysis tools than what it has now. For the purpose of demonstrating usage of ADLs for embedded system design and analysis, we selected AADL, but the method demonstrated here applies to EAST-ADL based design as well.

\begin{figure*}[!t]%trim=0cm 0cm 4cm 1cm, clip=true,
\centering
\includegraphics[trim=0cm 0cm 0cm 0cm, clip=true,width=0.88\textwidth]{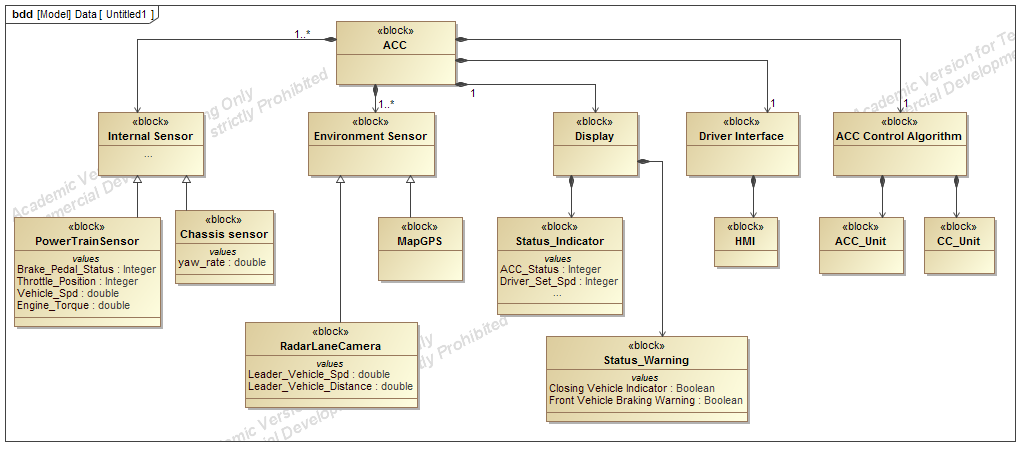}
\caption{Structure Diagram in MagicDraw shown above gives dependency and overall relationship of the different blocks.}%
\label{fig:SysML1}%
\end{figure*}

\begin{figure*}[!t]%trim=0cm 0cm 4cm 1cm, clip=true,
\centering
\subfloat[]{\includegraphics[trim=-0.4in 2.81in 5.6in 0in, clip=true, width=0.38\textwidth]{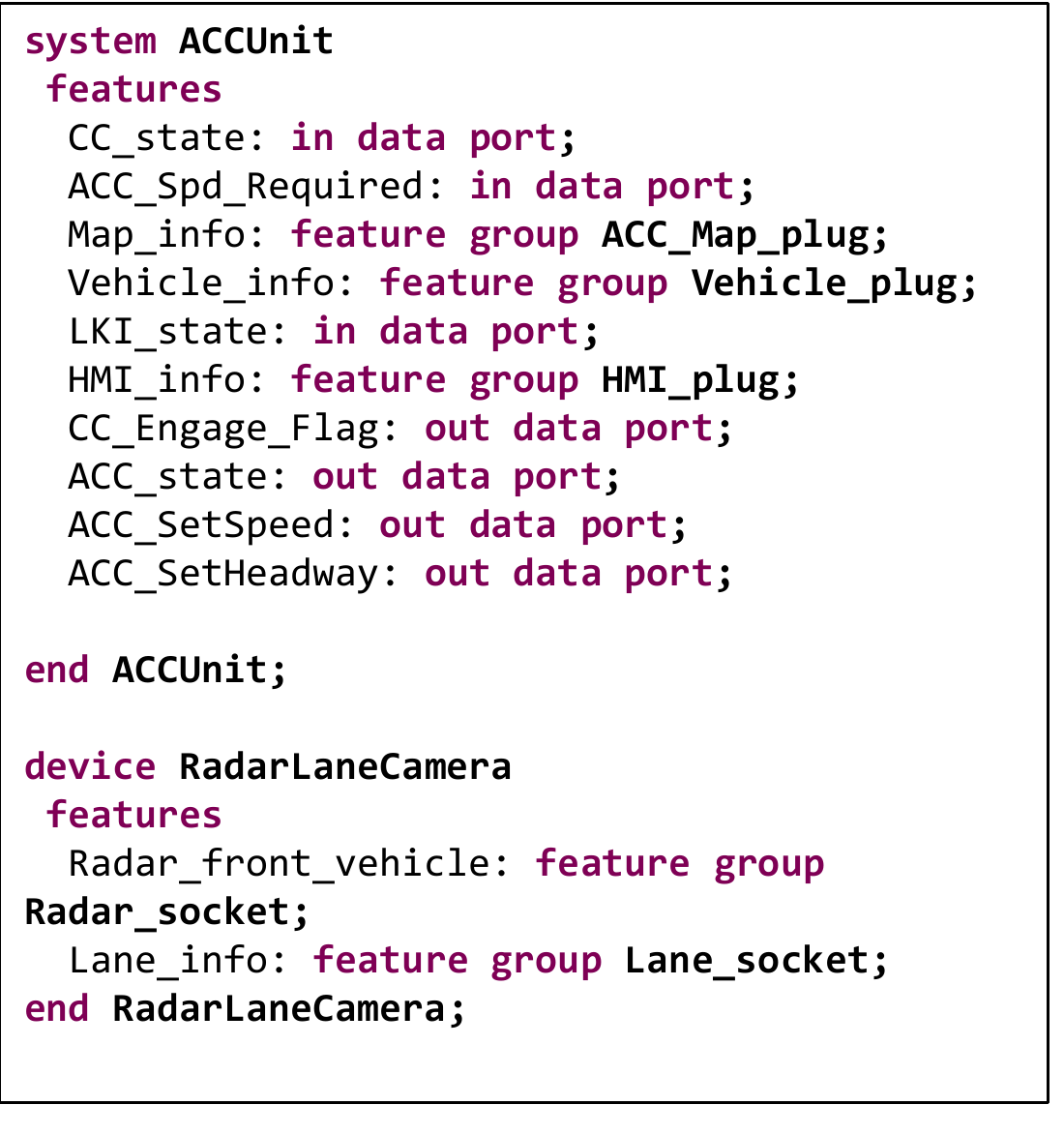}
 \label{fig:AADLGenerated1}
}
\hfill
\subfloat[]{
 \includegraphics[trim=4.5in 2.75in 0in 0in, clip=true, width=0.45\textwidth]{AADLModels1}
 \label{fig:AADLGenerated2}
}
\centering
\caption{(a) shows the AADL model type definition the new ACC controller unit and radar sensor. (b) is the corresponding AADL model of the overall system ACC that captures the internal connectivity between different blocks. Some connections and ports are omitted for concision. All the internal subsystems including ACCUnit have no implementation created at this phase. }%
\label{fig:AADLGenerated}%
\end{figure*}

AADL, as an architecture description language for embedded systems, gives the designer an opportunity to start from an abstract model and refine it progressively to a detailed high fidelity model. Such hierarchical design allows analysis and requirement verification for different stages of the design. The AADL model starts with definition of different component types of the system. These types are classified in standard categories including process, thread, thread group, data, subprogram, processor, memory, bus, device and system. The component types include interfaces or features to interact with the other parts of the system. The same component type can be implemented differently and therefore have completely different characteristics. Different categories have their standard properties associated and thus can be specified in their implementations, which can differ in internal structure and property such as scheduling protocol for processors. In the early phase, system engineers can use AADL to capture the high level model and describe just the interface level of the subcomponents. This abstract model can be transformed from SysML or other system design models commonly used in industry, because in the high level, the architecture information is very similar. In the second phase, controller designers use the AADL model as a guide to define behaviors of the models so the software components, such as a process in AADL model, will be refined to detailed function calls (subprograms), threads and thread groups. Hardware designers can perform design in parallel from the basic AADL model, such as proposing hardware platform choices and memory and power allocation based on requirements. The third phase starts with the integration of the hardware and software designs, then different properties and performance evaluations of the architecture are added to the model. This allows reliability, timing and error analysis to be performed in the earliest stage of design, reduces redesign cost and improves efficiency. 

The method we propose starts with requirements capturing and architecture modeling in the first phase. AADL coarse model can be generated from a system model such as System Modeling Language (SysML) or directly created based on the specifications. The second phase is software behavior modeling and hardware co-design, in other words, refining software and hardware architecture in detail in AADL and designing corresponding behavior in a simulation tool such as Simulink. The AADL model at this stage enables early verification such as error analysis, timing analysis and memory allocation. In the third phase, real-time simulation is performed based on the behavior model with timing properties associated to finalize hardware and software design parameters. Code generation and hardware implementation is performed at the last phase guided using the AADL model. The AADL model acts as a backbone to the whole design process. The first three phases of the work are demonstrated here emphasizing hardware-software co-design and analysis.

\section{ACC System Model}\label{sec:ACC}
The hardware software codesign, early verfication and analysis of AADL model is demonstrated on an automotive application, the Adaptive Cruise Control (ACC) algorithm.  When the radar and front camera sense the speed and distance of the front vehicle, the desired speed to be maintained for the host vehicle is computed by the algorithm. It also works as a cruise control unit when there is no leading vehicle in the range, i.e. maintains the speed set by the driver. 

The design starts with a very abstract model of the ACC system, which includes devices such as radar, speed sense and human inputs, and system blocks such as an existing cruise control unit and a new ACC controller block to be defined, shown in the SysML structure diagram Fig. \ref{fig:SysML1}. The equivalent AADL is transformed from the SysML model using model transformation as shown in Fig. \ref{fig:AADLGenerated}.  Hardware components such as the processor and memory are not defined. Such components are part of different subsystems which at this stage are not refined to include hardware or software subcomponents.  Overall system decomposition is shown on the right, defined by the subcomponents and the connections between different system interfaces.

\begin{figure*}[!t]%trim=0cm 0cm 4cm 1cm, clip=true,
\centering
\subfloat[]{\includegraphics[trim=0cm 3.35in 5.2in 0cm, clip=true, width=0.38\textwidth]{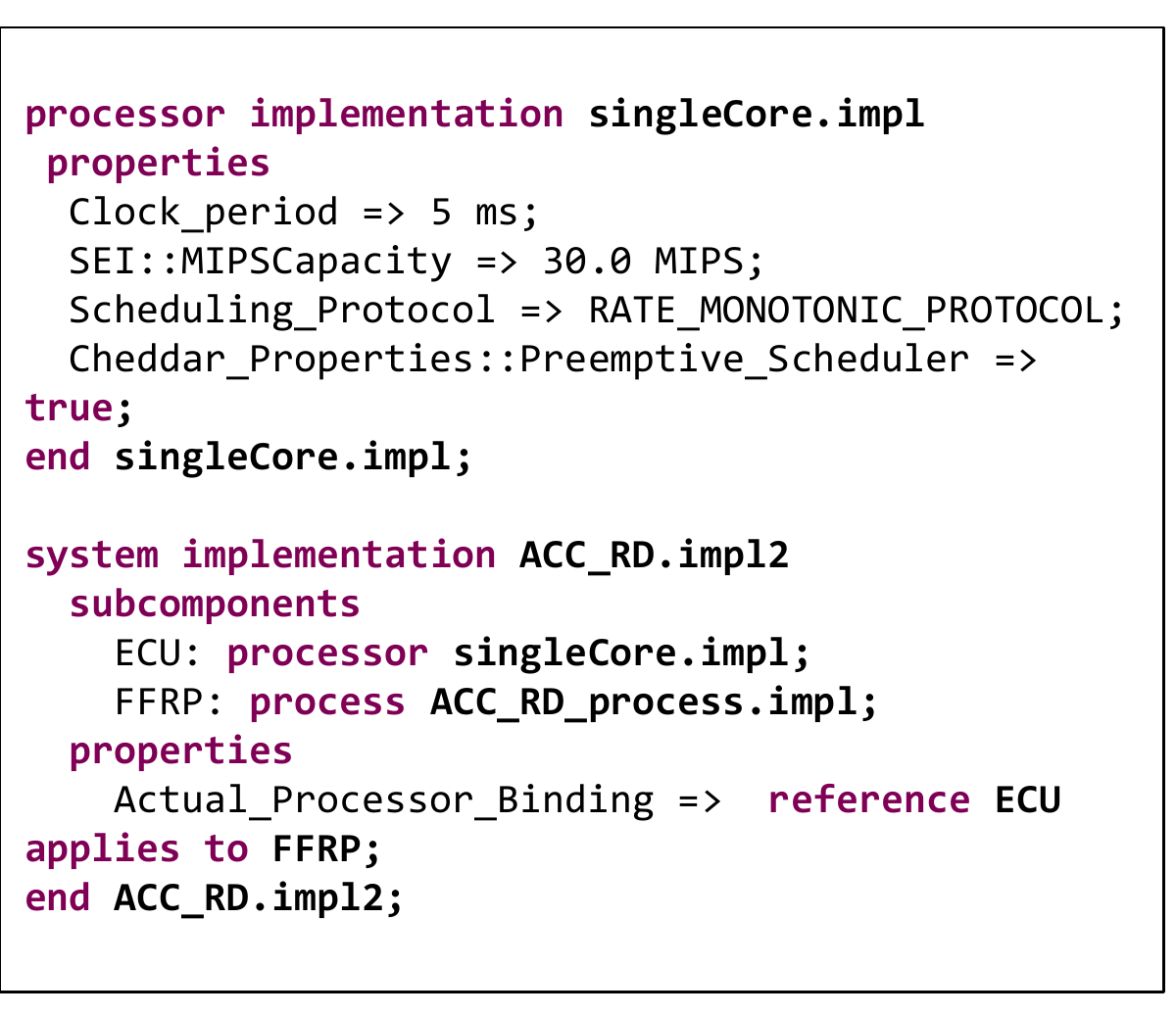}
 \label{fig:AADLCheddarSingle}
}
\hfill
\subfloat[]{
 \includegraphics[trim=5in 3.35in 0cm 0in, clip=true, width=0.38\textwidth]{AADLModels2}
 \label{fig:AADLCheddarDual}
}
\centering
\caption{(a) shows the single core configuration of ACC system, whereas (b) shows the dual-core configuration. The software component FFRP process is defined as an instance of type ACC\_RD\_process.impl, which refers to what the software group is developing at the same time. The AADL model ACC\_RD.impl is an implementation of an extension of system type ACCUnit in Fig. \ref{fig:AADLGenerated}. }%
\label{fig:AADLHardware}%
\end{figure*}

From a controller performance point of view, the ACC algorithm is required to keep the relative distance between leading and host vehicle to be at least 90\% of the driver-selected headway, which specifies the distance in time for the host vehicle to reach the leader. From the hardware implementation point of view, the overall ACC algorithm needs to be executed on either an existing single core processor or a dual-core processor with the same computational power. Because of the close relationship between hardware and software requirements, co-design is necessary to fulfill requirements from both areas. Thus the goal is to select the hardware architecture that has tolerable controller performance degradation and lower cost.

\section{Refined AADL for Analysis and Hardware-Software Co-design} \label{sec:codesign}
The coarse model in the previous step is refined further here. This section consists of two aspects, one is to capture the AADL hardware requirements, the second is to decompose the AADL software model further to the threads level. Two aspects are developed in parallel, independent of one another and integrated at the end. 

\begin{figure*}[!t]%trim=0cm 0cm 4cm 1cm, clip=true,
\centering
\subfloat[]{
 \includegraphics[trim=0in 3.76in 5.05in 0in, clip=true,width=0.38\textwidth]{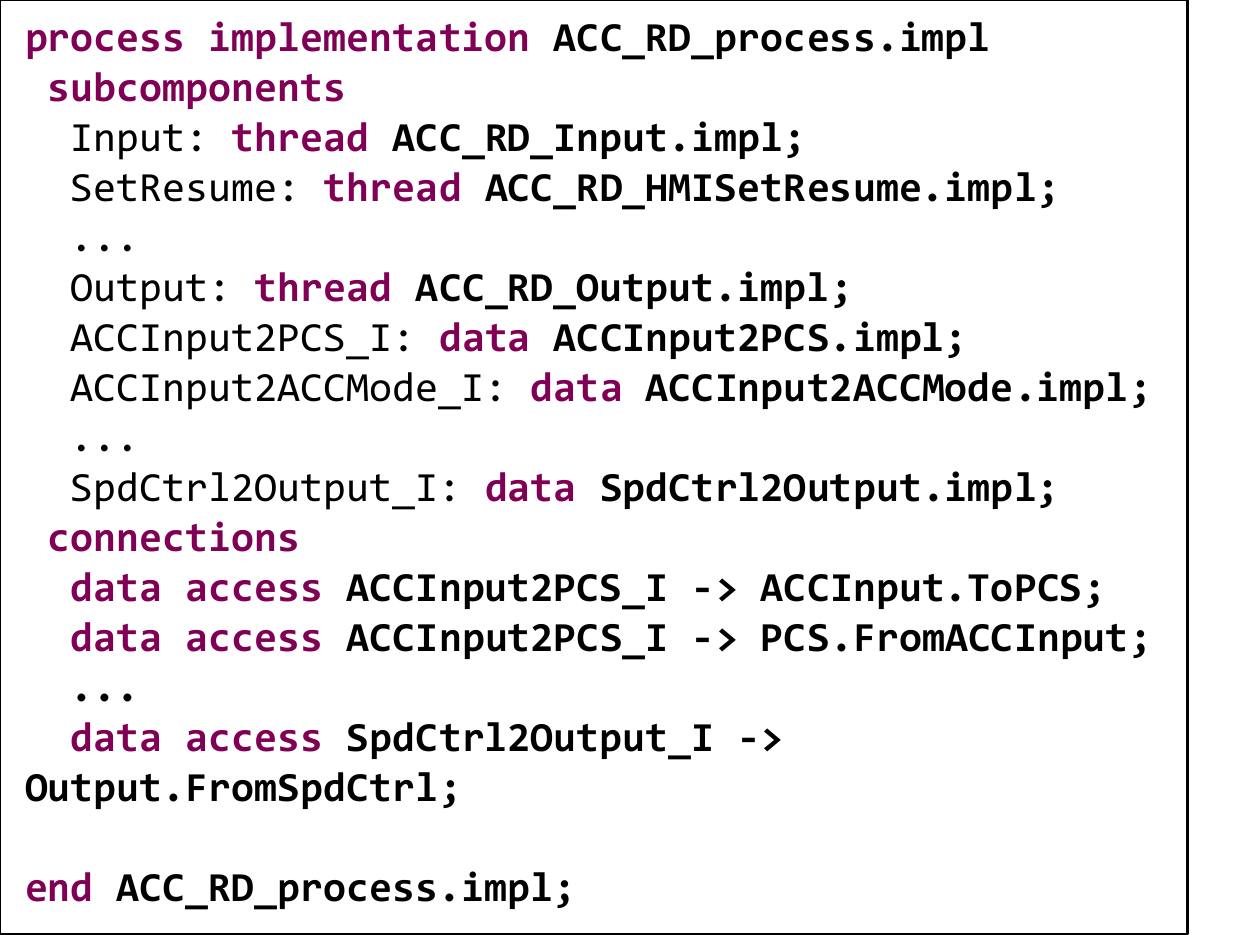}
 \label{fig:AADLCheddar1}
 }
\hfill
\subfloat[]{
 \includegraphics[trim=5in 3.76in 0in 0in, clip=true,width=0.38\textwidth]{AADLModels3}
 \label{fig:AADLCheddar2}
}
\caption{(a) shows ACC process implementation model with internal structures such as threads and data. (b) shows one of ACC threads and its implementations with associated Cheddar timing property. }
\label{fig:AADLCheddar}
\end{figure*}

\begin{figure*}[!t]%trim=0cm 0cm 4cm 1cm, clip=true,
\centering
\includegraphics[trim=0cm 0cm 10cm 0cm, clip=true,width=0.86\textwidth]{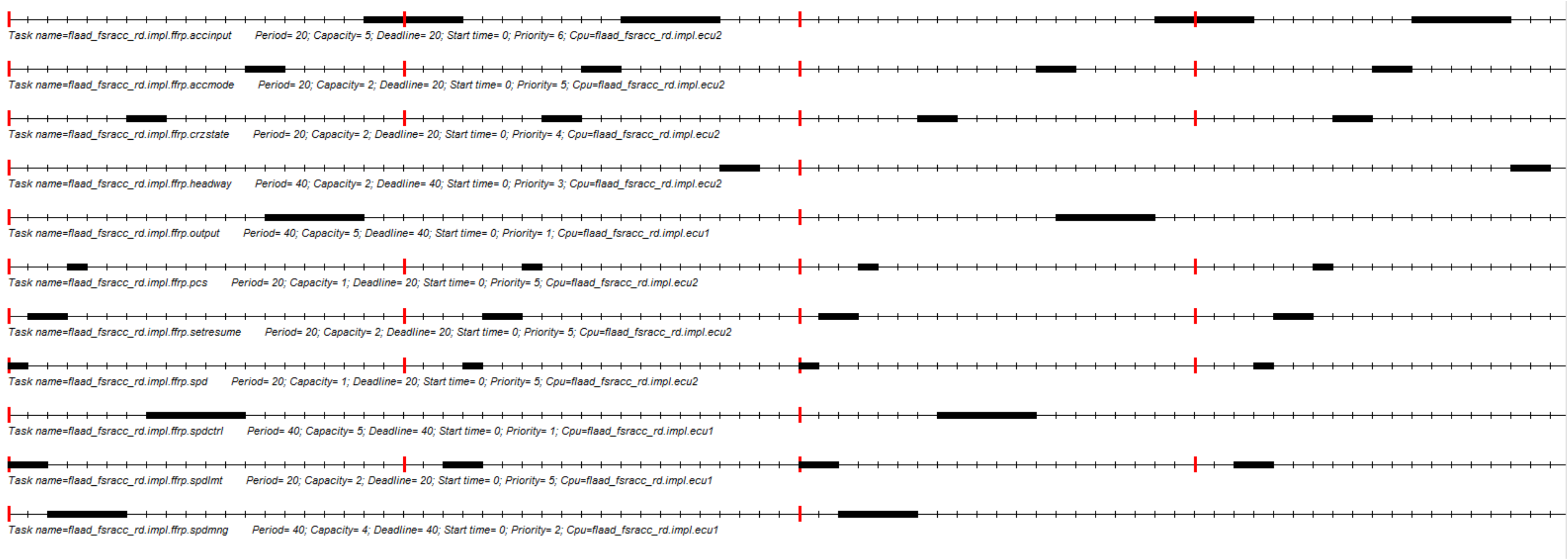}
\caption{Cheddar simulation results of shared data access for dual-core setup.}%
\label{fig:CheddarScheduler}%
\end{figure*}

Since the hardware requirements are not finalized to a single core or multicore configuration, two hardware configurations are proposed in AADL (Fig. \ref{fig:AADLHardware}). At the same time, the software group developed Simulink blocks based on the architecture defined in the AADL models. Then based on the Simulink blocks, AADL subsystems are decomposed to threads, which enables analysis tools to link with the AADL model. Timing analysis is emphasized here to help the selection of the hardware configuration. Since both hardware configurations use similar computation power, every thread is tested and assigned an estimated computation time. Given the sensor and signal processor system, the maximum sampling rate for input inquiry thread is also assigned. All of this information can be captured in AADL using the predefined properties of threads and processes. In order to fully facilitate the timing and scheduling analysis enabled by AADL, Cheddar is chosen to be the external tool to analyze the XML model generated by the AADL model. The AADL model can be imported to Cheddar externally. By specifying the appropriate scheduling protocols and data sharing protocols, Cheddar can perform the scheduling feasibility analysis and simulation. 
Given the execution time and maximum sampling rate of sensor input threads, the dual-core and single core configurations are evaluated with different thread sampling periods and scheduling protocols. A concise AADL model is shown in Fig. \ref{fig:AADLCheddar}. The generated scheduler of different threads for a dual-core configuration is shown in Fig. \ref{fig:CheddarScheduler}, based on Rate Monotonic scheduling and Best Fit partition allocation. The feasibility test returns the utilization of ECU1 to be 45\% and ECU2 to be 70\%. As can be seen, the ECU2 is almost fully utilized, and the period is 40ms for this case. A single core configuration requires the period to be at least 70ms for utilization less than 71\% for 10 threads based on Liu \& Layland \cite{RMLiu}. We chose 80ms for single core and 40ms for dual-core as the reference period to perform the controller evaluation in the next section.

\section{AADL Real-Time System Simulation and Hardware Selection}

The Cheddar tool is used for sampling period selection and feasibility testing, whereas Simulink Truetime, a real-time simulation toolbox for Simulink \cite{cervin2003does}, is used to evaluate the controller performance degradation. Truetime extends Simulink functionality on timing analysis by adding data passing and timing delays between threads and processes. The Truetime models are constructed based on the controller designed and the AADL models. It is possible to translate the AADL model to Truetime script directly using metamodel transformation, since they captures the same information. It is not practiced here since the models are rather simple. It is a part of future work to link AADL and Truetime, when such method is applied to larger system. For the single core case, the task is given fixed priority and tasks/threads are running consequentially based on priority. For the dual-core case, simulation is set up so that each core is synchronized with the other with the same priority setting, e.g. when one of the cores is writing to some memory locations, the other core has to wait until the first finishes before reading those particular memory locations. Physical plants, i.e. the host car and leading car, are co-simulated in the CarSim environment with Simulink controller blocks.

The objective here is to simulate ACC under the same road scenario for single and dual-core configurations. The controller performance has to satisfy the requirement of headway and the speed profile should be smooth and stable for the host vehicle. Different sampling periods are also examined to evaluate timing effects on controller performance. All results are compared to a reference model without timing effects. The final goal is to select the best hardware setup and sampling period for the algorithm.

\begin{figure*}[t]%trim=0cm 0cm 4cm 1cm, clip=true,
\centering
\subfloat[Control performance for period 40ms.]{
 \includegraphics[trim=0cm 0cm 0cm 0cm, clip=true, width=0.45\textwidth]{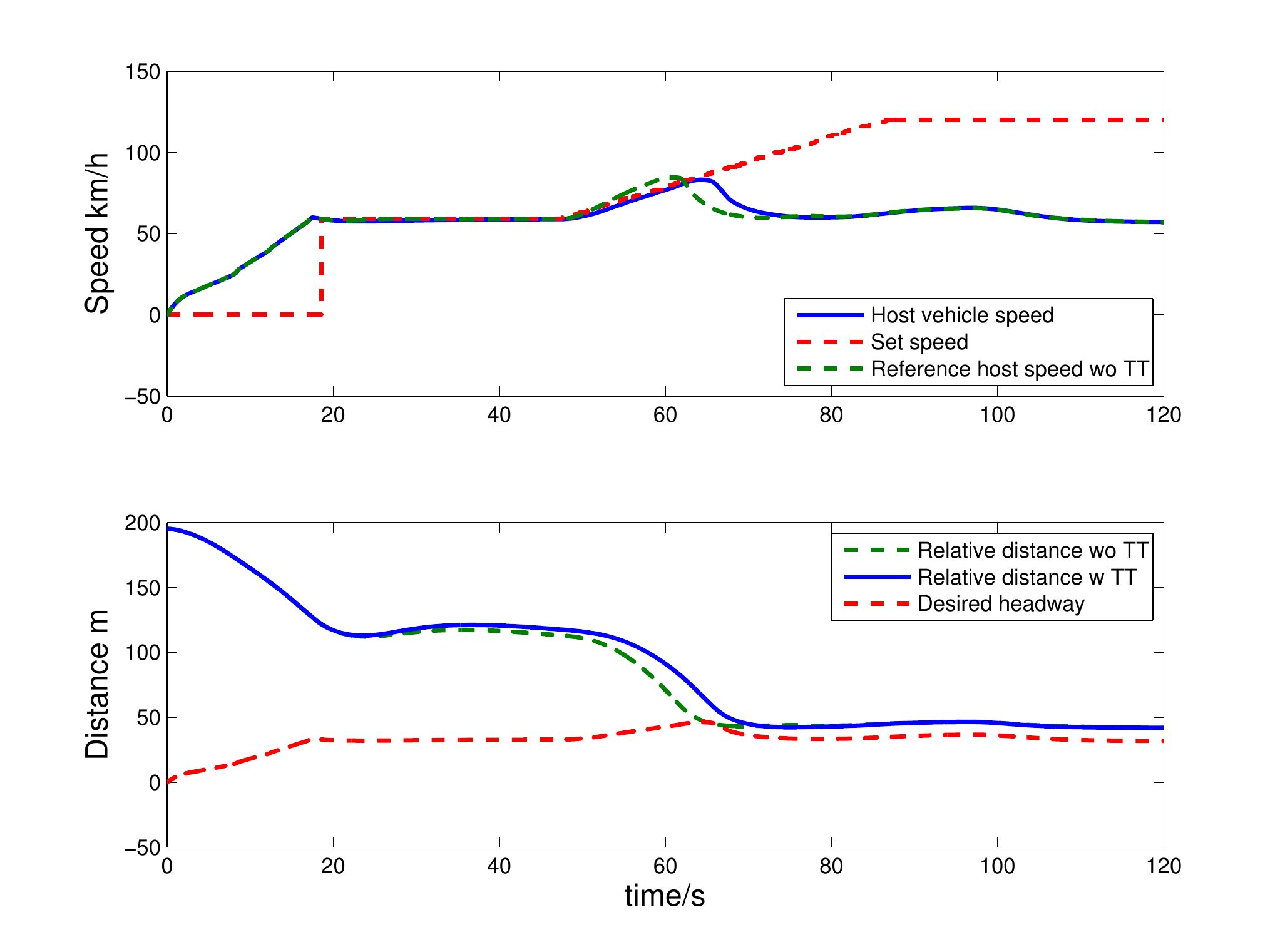}
 \label{fig:normal_single_20ms_cp}
}
\subfloat[Schedule for period 40ms.]{
 \includegraphics[trim=0cm 0cm 0cm 0cm, clip=true, width=0.38\textwidth]{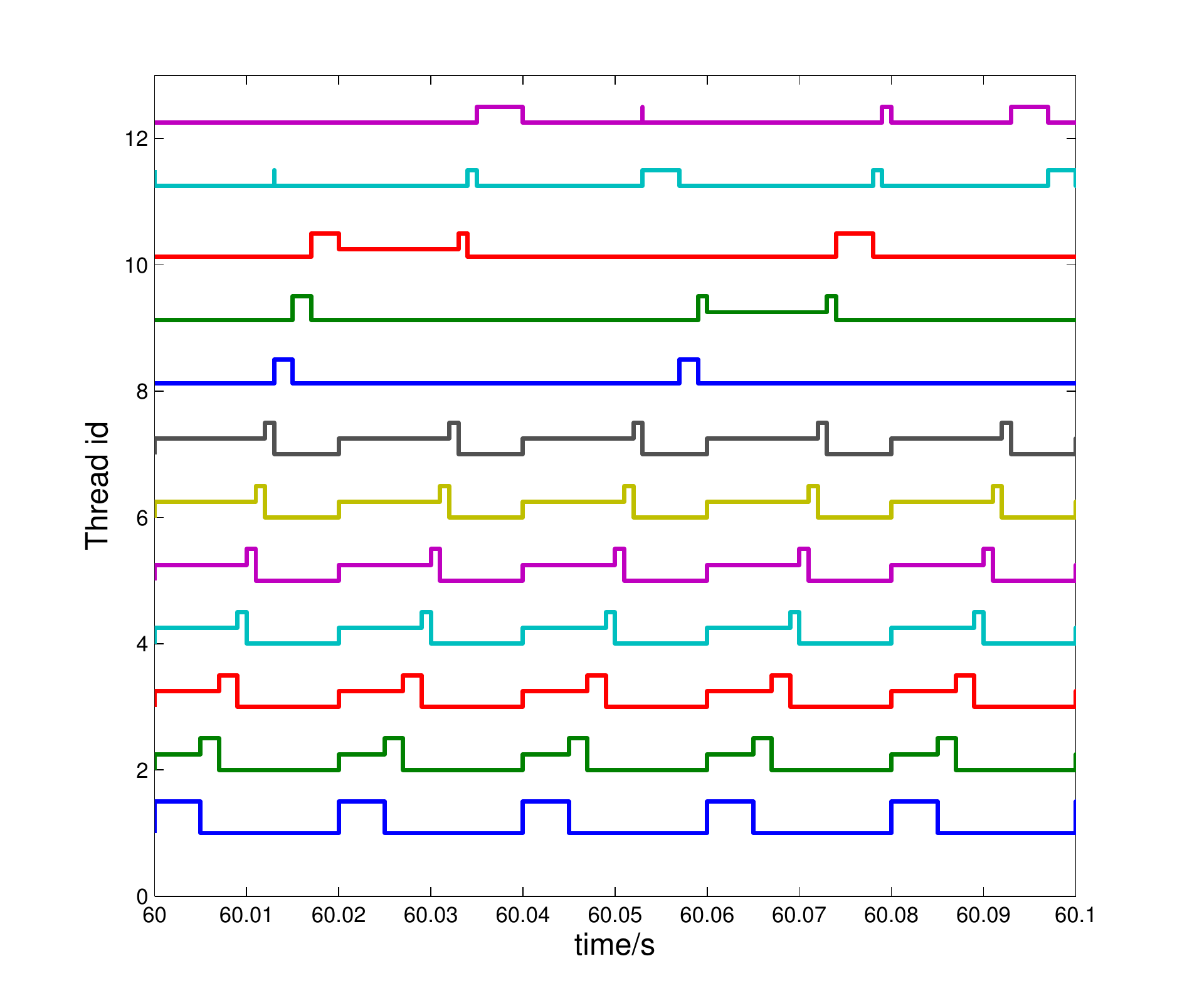}
 \label{fig:normal_single_20ms_sch}
}\,
\subfloat[Control performance for period 80ms.]{
 \includegraphics[trim=0cm 0cm 0cm 0cm, clip=true, width=0.45\textwidth]{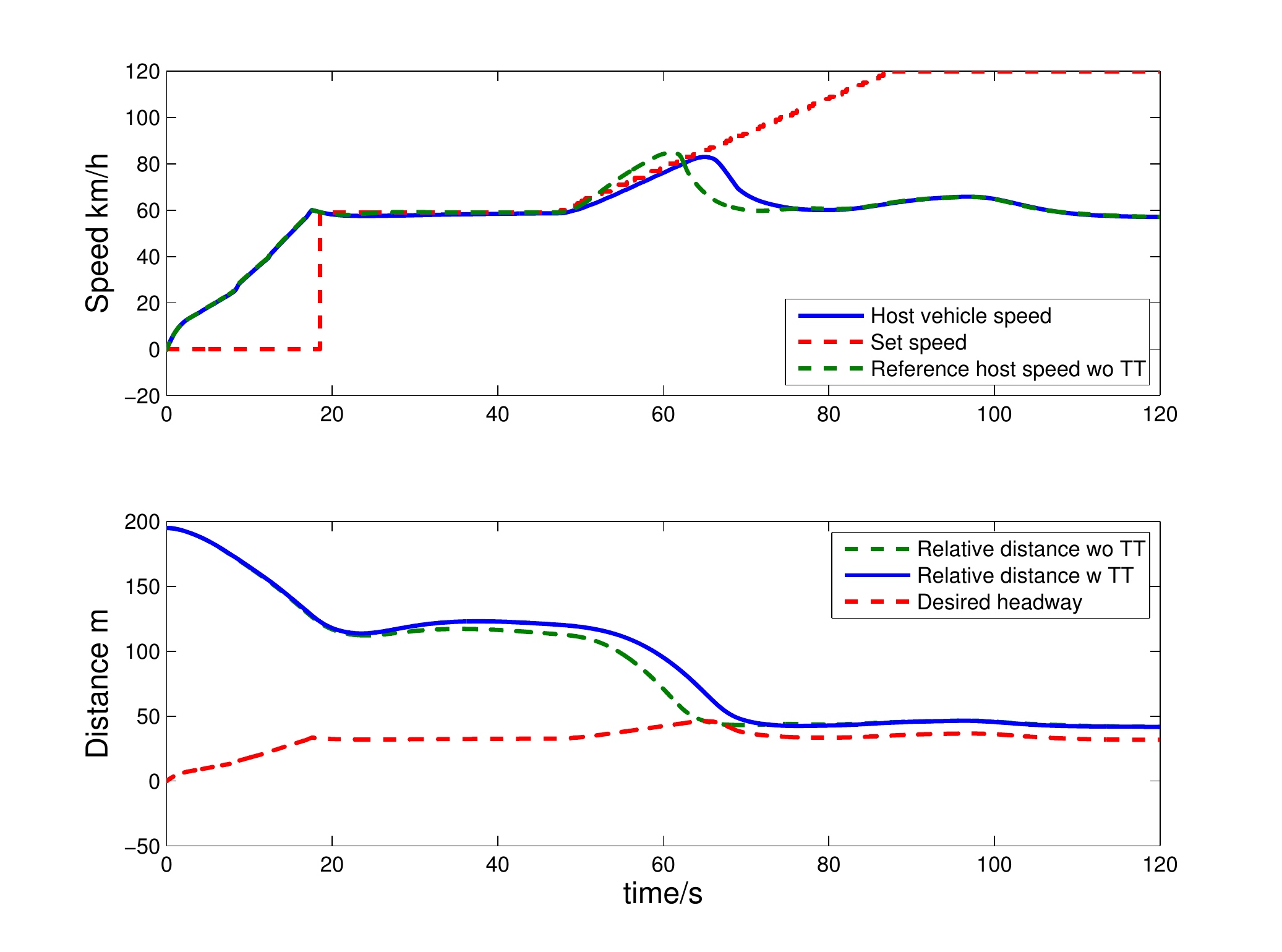}
 \label{fig:normal_single_40ms_cp}
}
\subfloat[Schedule for period 80ms.]{
 \includegraphics[trim=0cm 0cm 0cm 0cm, clip=true, width=0.38\textwidth]{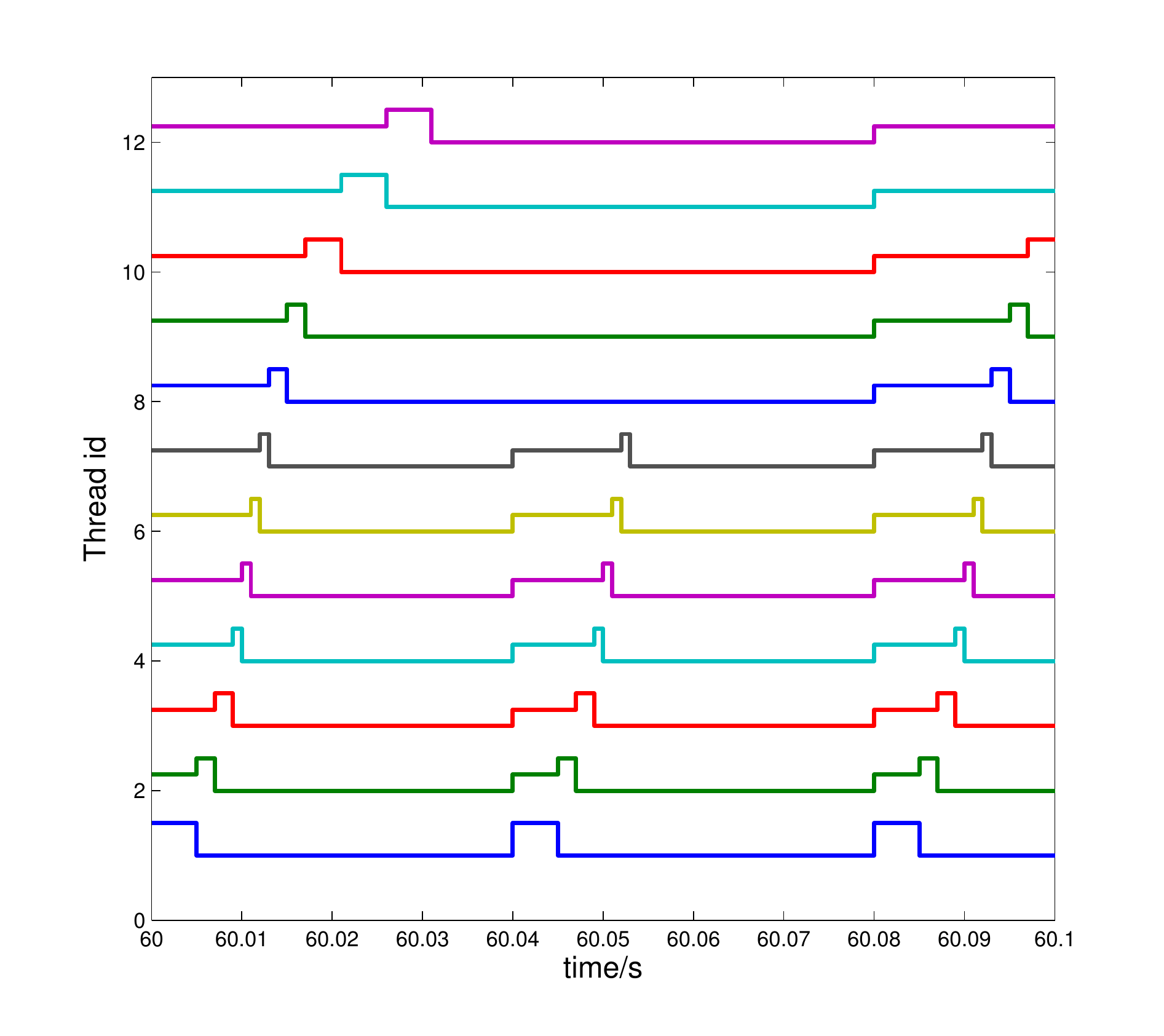}
 \label{fig:normal_single_40ms_sch}
}\,
\caption{Control performance of single core case is shown in the left column. The top subplot of each figure shows the speed profile of the following/host vehicle. The bottom subplot of the figure shows the relative distance between the cars. The scheduler outputs of the corresponding cases are shown in the right column.}
\label{fig:truetime_single1}
\end{figure*}
	
	\subsection{Single Core Case}
		
	Based on Cheddar analysis, a sampling rate greater than 70ms is needed. For simplicity, 80ms sampling is used as reference model. The other sampling rate to be compared is 40ms when threads are not schedulable and 160ms for downsampling. One reason is to evaluate the behavior of the algorithm under a fault event when it is oversampling or downsampling. The other reason is to estimate the largest possible sampling rate for this algorithm to work and evaluate the possibility of executing it in parallel with other related algorithms.
	
	For all cases, the leading vehicle speed profile is the same with a fixed speed around 60km/h. The ``set speed increase'' button is pressed on the host vehicle for 45s until the set speed reaches 120km/h. The host vehicle decreases the speed after around 60s due to the decrease in relative distance from the front vehicle. The results for 40ms and 80ms sampling are shown in Fig. \ref{fig:truetime_single1}. The 160ms sampling is similar to the 80ms sampling result. In the relative distance plots, the green dashed line is a reference profile where Truetime is not used. The red line is the lower boundary based on the headway specification. The relative distance as shown for both cases is higher than the requirement boundary determined by the headway distance.
	
	As can be seen in the scheduler output with a 40ms period, several threads cannot be executed within the deadline. This is expected based on scheduling analysis. However, there is no large controller performance degradation, which means for a single core algorithm, it does provide robustness to timing effects. The degradation seen in the 80ms result is caused by the fact that the algorithm is initially designed for a 40ms controller. The conclusion is that for the single core configuration, 80ms sampling is the best among all the tested sampling rates. The degradation of controller performance is tolerable. 
	
	\subsection {Dual-core Case}
	
\begin{figure*}[t]%trim=0cm 0cm 4cm 1cm, clip=true,
\centering
\subfloat[Control performance for period 20ms.]{
 \includegraphics[trim=0cm 0cm 0cm 0cm, clip=true, width=0.45\textwidth]{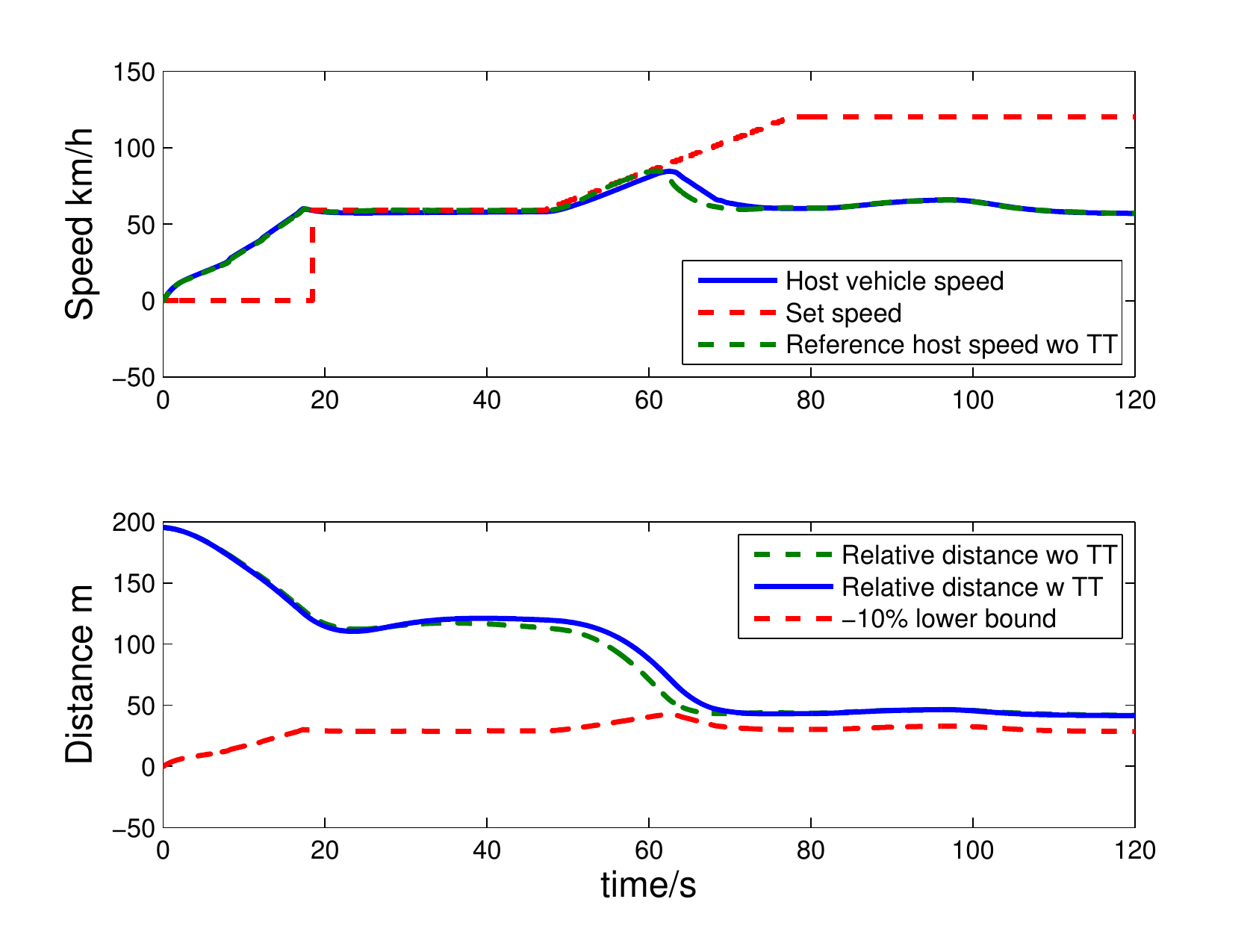}
 \label{fig:truetime20ms_cp}
}
\subfloat[Schedule for period 20ms.]{
 \includegraphics[trim=0cm 0cm 0cm 0cm, clip=true, width=0.4\textwidth]{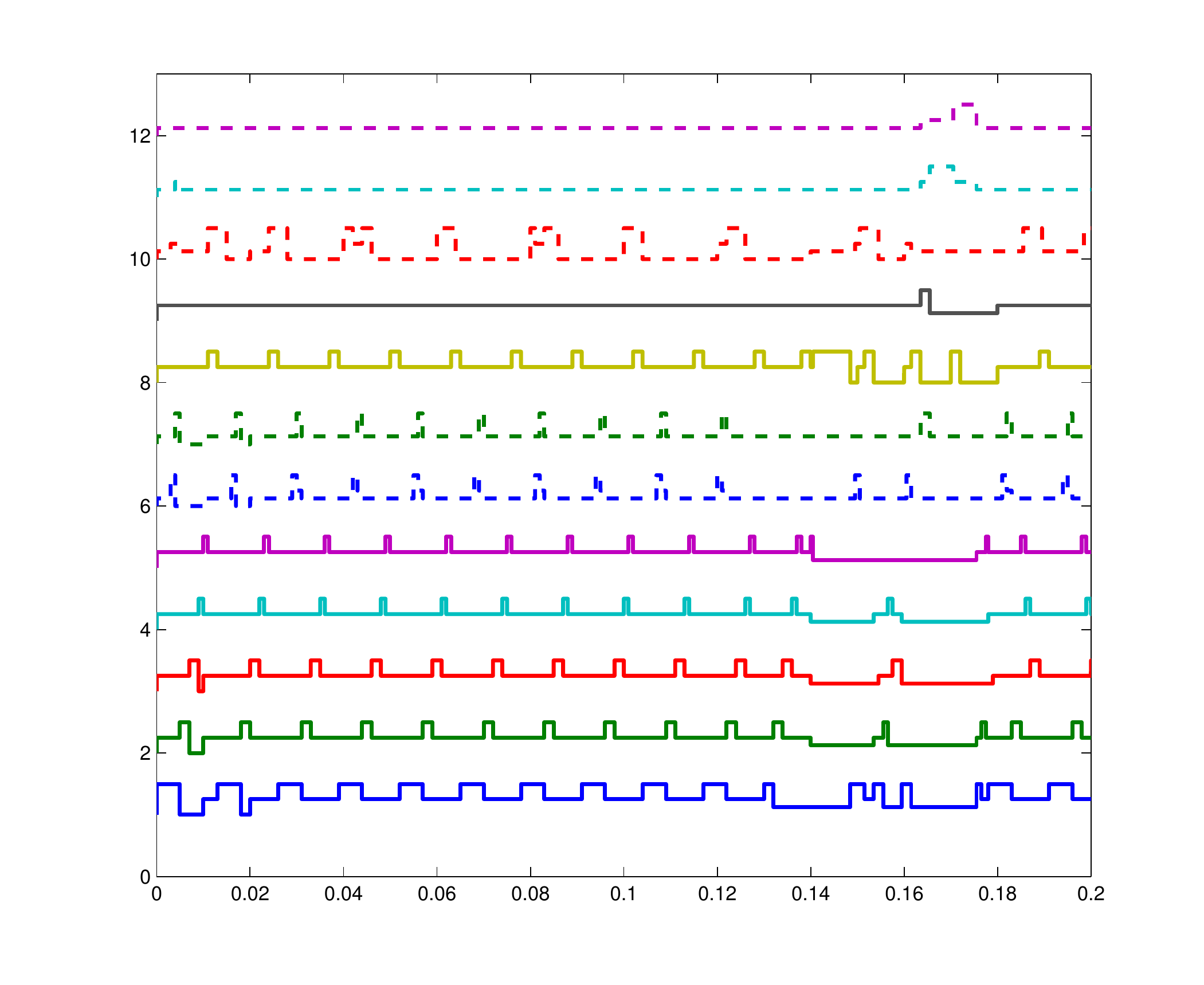}
 \label{fig:truetime20ms_sch}
}\,
\subfloat[Control performance for period 40ms.]{
 \includegraphics[trim=0cm 0cm 0cm 0cm, clip=true, width=0.45\textwidth]{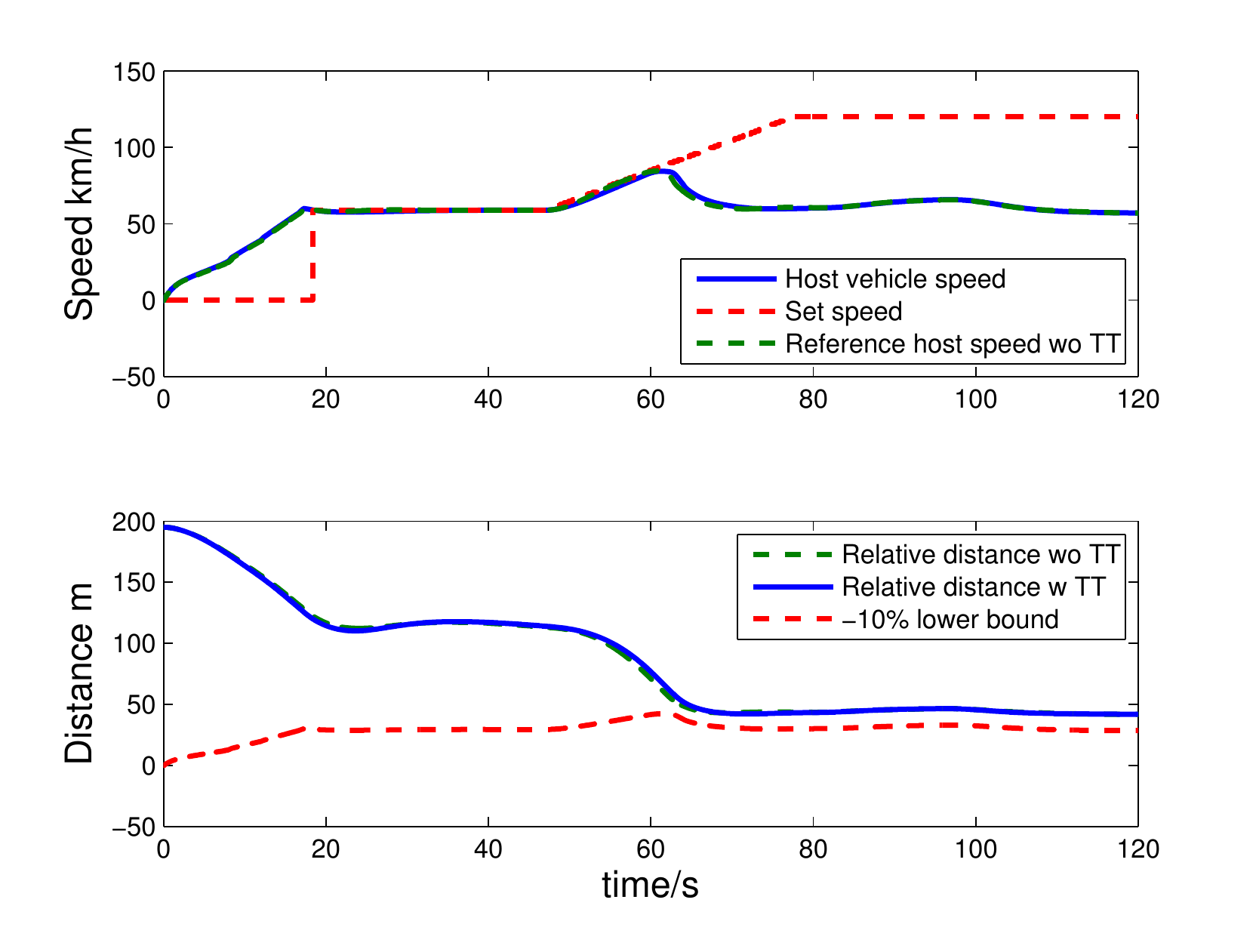}
 \label{fig:truetime40ms_cp}
}
\subfloat[Schedule for period 40ms.]{
 \includegraphics[trim=0cm 0cm 0cm 0cm, clip=true, width=0.4\textwidth]{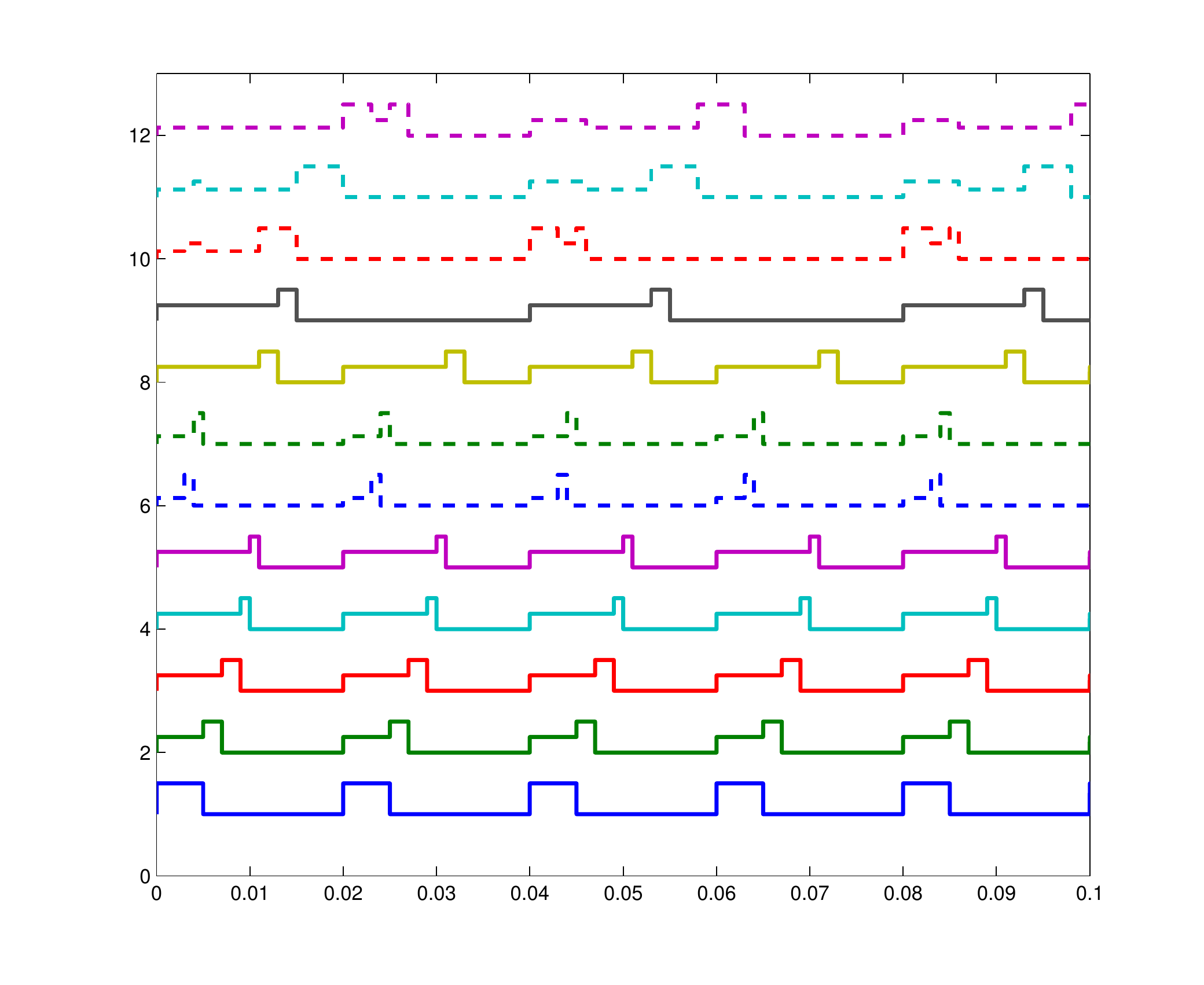}
 \label{fig:truetime40ms_sch}
}\,
\caption{Control performance of dual-core cases are shown in the left column. The top subplot of each figure shows the speed profile of the following/host vehicle. The bottom subplot of the figure shows the relative distance between the cars. The scheduled outputs of the corresponding cases are shown in the right column.}
\label{fig:truetime1}
\end{figure*}
	
	Similar to the single core case, sampling periods of 20ms, 40ms, 80ms with the same execution time for each thread are chosen to observe the performance degradation caused by oversampling and downsampling. The simulation setup is the same as the regular one in single core case. Controller performance of 20ms and 80ms is very similar, so the 80ms one is not shown. As can be observed from Fig. \ref{fig:truetime1}, the relative distance is always higher than the lower boundary determined by headway distance. As expected, controller performance for 40ms is almost the same as the case without timing effects since the algorithm is designed for 40ms sampling. On the scheduler side, the dotted lines indicate the scheduler for ECU2, while the solid lines indicate the scheduler for ECU1. Clearly the system behaves abnormally for 20ms sampling which is expected based on scheduling analysis. For 40ms, the scheduler output is desirable. Thread 6,7 in ECU2 are executed in parallel and synchronized with the behavior of thread 1 in ECU1.  The conclusion is that for a dual-core setup, 40ms sampling is the best among all the tested sampling rates. 
		
		For common implementation when the ECU is dedicated to the ACC algorithm only, the single core configuration has similar performance compared to dual-core configuration in terms of the headway requirement. Thus single core with 80ms sampling is preferred as it is easier to implement and lower in price.

\section{Conclusion}
Based on the ACC experiment we conclude that hardware-software co-design using AADL is well suited for early verification and requirement refinement in industry practices. Architecture information is important especially in the early design phase to capture requirements and perform associated analysis. The real-time simulation is necessary for detailed controller performance evaluation. Our method is well suited for model based development that is required in industrial embedded system design. Future work is to apply such methods to larger systems and perform other analysis attractive to industry, such as error analysis and power allocation analysis as enabled by the AADL model.

%--------------------------------------------------------------------------------
% Acknowledgements
\acks
The authors would like to thanks Joseph D'Ambrosio, Lei~Rao, Rami~I~Debouk from General Motors, Electrical and Controls System Lab for the inspiring discussion on AADL and its tool OSATE. 

This research is supported by the National Science Foundation (NSF) under grant award CNS-1035655.
%\IEEEtriggeratref{11}


\bibliographystyle{plain}
\bibliography{bib}
%\begin{thebibliography}{1}
  %0.5em minus 0.4em\relax Harlow, England: Addison-Wesley, 1999.
%
%\end{thebibliography}

\end{document}